\documentclass[utf8
               ]{jetp} 
\pdfoutput=1
\twocolumn 

\usepackage{mathrsfs}
\usepackage[percent]{overpic}

\begin{document}
{
\English

\title{ANISOTROPIC FEATURES OF THE TWO-DIMENSIONAL HYDROGEN ATOM IN A MAGNETIC FIELD}

\setaffiliation1{Bogoliubov Laboratory of Theoretical Physics of JINR  \\ 141980, Dubna, Russia} %
\setaffiliation2{Department of Fundamental and Applied Problems of Microworld Physics, State University “Dubna"\\ 141980, Dubna, Russia}

\setauthor{E.~A.}{Koval}{12}
\email{e-cov@yandex.ru}

\setauthor{O.~A.}{Koval}{1}
\email{kov.oksana20@gmail.com}

\rtitle{Anisotropic features of the two-dimensional Hydrogen in a magnetic field}
\rauthor{E.~A.~Koval, O.~A.~Koval}

\abstract{
The aim of the current work 
is the numerical research of
the anisotropic characteristics of the two-dimensional hydrogen atom induced by a magnetic field. The ground state energy (GSE) of the two-dimensional hydrogen atom and the corresponding wave function have been  numerically calculated in the Born-Oppenheimer approximation and with taking into account the finite mass of the proton. The non-linear dependence of GSE on the angle $\alpha$ between the magnetic field vector and the normal to the plane of electron motion in a wide range of magnetic field strength has been found. The effect of a significant reduction of GSE (up to 1.9-fold) is observed with increasing the angle $\alpha$ up to $90^{\circ}$.

}

\maketitle

\section{INTRODUCTION}

The interest in two-dimensional (2D) systems has been maintained due to a wide range of effects emerging in them: Berezinskii – Kosterlitz – Thouless transition ~\cite{
KosterlitzThouless_1973}, fractional quantum Hall effect in an tilted external magnetic field ~\cite{Engel_1992,Eisenstein_1992}, superconductivity in quasi-2D organic conductors induced by a magnetic field ~\cite{Kobayashi_2001}, prediction ~\cite{Sofo_2007} and discovery ~\cite{Novoselov_2009} of graphane, which is a quasi-2D monolayer of graphene bound to atomic hydrogen, etc.

Initially, the 2D model of the hydrogen atom was investigated 
within
purely theoretical considerations ~\cite{Zaslow_1967,Huang_1979,Hassoun_1981,Yang_1991}, but it was also applied to describe highly anisotropic three-dimensional crystals ~\cite{Kohn_1955}. With the development of experimental methods for creation of low-dimensional systems and new prospects for development of semiconductor devices, the 2D hydrogen model was used to describe the effect of a charged impurity in 2D systems ~\cite{Chen_1991,Villalba_1996,Soylu_2007} and effective interaction in the exciton electron-hole pair, the motion of which is limited by the plane, in semiconductor 2D heterostructures ~\cite{Portnoi_2002}. A number of studies investigated the internal symmetries of the model and the reasons for accidental degeneracy occurring in the three-dimensional (3D) case as well ~\cite{Portnoi_2002,Cisneros_1969,Robnik_1981}.

The influence of the external magnetic field, which is 
perpendicular
to the plane of electron motion, on the spectrum of 2D hydrogen was investigated by means of a two-point Pade approximation ~\cite{MacDonald_1986}, method of asymptotic iterations ~\cite{Soylu_2006}, variational approach ~\cite{Turbiner_2014,Turbiner_2015}, as well as analytically for individual values of the magnetic field \cite{Taut_1995}. Studies of hydrogen in strong magnetic fields ~\cite{Robnik_2003, Vinitsky2005, Vinitsky2007} are associated with the astrophysical applications: the magnitude of the magnetic field in dwarf stars reaches $10^2$ -- $10^5$ T, whereas in neutron stars~--- $10^7$ -- $10^9$ T~\cite{Ruder_2012}.

The aim of the current work is the numerical investigation of the anisotropic properties of the two-dimensional hydrogen atom induced by a magnetic field. In contrast to the previous works ~\cite{MacDonald_1986,Soylu_2006,Turbiner_2014,Turbiner_2015,Taut_1995,Robnik_2003}, we study the dependence of the spectrum and the wave functions of the system 
on arbitrary directions of the magnetic field forming an angle $\alpha$ with the normal to the plane of electron motion (see Fig.  ~\ref{fig1}). It should be noted that, although the electron in the 2D hydrogen atom moves in a plane, the electromagnetic fields, angular momentum and other values are not limited to the plane.

To discretize the Hamiltonian matrix, we used the wave function expansion ~\cite{Melezhik_1991,Koval_2014} based on the discrete variable method, seven-point finite-difference approximation, whereas to solve the eigenvalue problem~--- the shifted  inverse iteration method ~\cite{Kalitkin_2011}. In each iteration, the matrix equation was solved by matrix modification of the sweep algorithm ~\cite{Gelfand_2000}.

In Chapter \ref{sectionAlgorithm} of this paper we generalize the problem of the 2D hydrogen atom bound states in a magnetic field for the case when the magnetic field is tilted to an angle $\alpha$ with respect to the
 normal to the plane of electron motion. We also describe an algorithm for its numerical solution. The results and their brief analysis are presented in Chapter \ref{Results}. Chapter \ref{Conclusions} gives the main conclusions. 

All values in the article are given in atomic units: $\hbar =m_e =e=1$.

\section{BOUND STATES OF THE 2D HYDROGEN ATOM IN AN EXTERNAL MAGNETIC FIELD}
\label{sectionAlgorithm}

The Hamiltonian of the 2D hydrogen atom in a uniform magnetic field $\boldsymbol{ \mathrm B}$ in the polar coordinates $\boldsymbol{\mathrm \rho}=(\rho,\phi)$ has the form \cite{Turbiner_2014}:
\begin{equation}
\label{eq1}
\mathscr{H}=\frac{\left( {\boldsymbol{\mathrm P}-2 \boldsymbol{\mathrm A}_{\rho}} \right)^2}{2(m_p + m_e) }+\frac{\left( {\boldsymbol{\mathrm p}- (\mu_e-\mu_p) \boldsymbol{\mathrm A}_{\rho}} \right)^2}{2m_r }-\frac{1}{\rho},
\end{equation}
where $m_p$~---mass of proton, $m_e$~--- mass of electron (corresponding reduced masses~-- $\mu_p=\frac{m_p}{m_p+m_e}$, $\mu_e=\frac{m_e}{m_p+m_e}$), $\boldsymbol{\mathrm P}$~--- total momentum, and $\boldsymbol{\mathrm p}$~--- relative momentum of the system; $m_r=\frac{m_p m_e}{m_p+m_e}$~--- reduced mass of the system. 
We used symmetric gauge for the vector-potential ${\boldsymbol{\mathrm A}_{\rho}=\tfrac{1}{2}\left[ {\boldsymbol{\mathrm B}\times \boldsymbol{\mathrm \rho} } \right]}$, where $\boldsymbol{\mathrm \rho}$ ~--- relative coordinate.

Let us choose a coordinate system considering its convenience and occurrence of axial symmetry in the two-dimensional hydrogen atom: the $XY$ plane coincides with the plane of electron motion, while the magnetic field $\boldsymbol{\mathrm B}=B 
\sin (\alpha )\boldsymbol{\mathrm i} +B \cos (\alpha )\boldsymbol{\mathrm k} $ lies in the $XZ$ plane (see Fig. \ref{fig1}).

Similarly to ~\cite{Turbiner_2014}, we considered a system at rest ($\boldsymbol{\mathrm P}=0$): in this case, the motion of the mass center in (\ref{eq1}) is separated from the relative motion. Using the representation of the wave function in the form:
\begin{equation}
\label{eqTurbinerAnsatz}
\Psi(\boldsymbol{\mathrm R},\boldsymbol{\mathrm \rho}) = 
e^{i\boldsymbol{\mathrm  P}\cdot\boldsymbol{\mathrm  R}}\Psi(\boldsymbol{\mathrm \rho}),
\end{equation}
where $\boldsymbol{\mathrm R}$ ~--- coordinate of the mass center, we will obtain the Hamiltonian of relative motion \cite{Turbiner_2014}:
\begin{equation}
\label{eqRelativeMotionMAIN}
H = 
\frac{\boldsymbol{\mathrm p}^2 + 2  (\mu_p-\mu_e) ( \boldsymbol{\mathrm A}_{\rho} \cdot \boldsymbol{\mathrm p})+ \boldsymbol{\mathrm A}_{\rho}^2}{2m_r}-\frac{1}{\rho}. 
\end{equation}
In ~(\ref{eqRelativeMotionMAIN}), the 2D kinetic energy operator 
\begin{equation}
\label{eq2DKineticPart}
\frac{\boldsymbol{\mathrm p}^2}{2m_r } = - \frac{1}{2m_r}\left( {\frac{1}{\rho
 }\frac{\partial }{\partial \rho }\left( {\rho \frac{\partial }{\partial \rho
 }} \right)+\frac{1}{\rho ^2}h^{(0)}(\phi )} \right)
\end{equation}
includes an angular part having a simple form in the polar coordinates $h^{(0)}(\phi )=\frac{\partial ^2}{\partial \phi ^2}$. The linear term
\begin{equation}
\label{eqLinearMagneticTermAnisotropy}
( \boldsymbol{\mathrm A}_{\rho} \cdot \boldsymbol{\mathrm p}) =  \tfrac{1}{2} (\boldsymbol{\mathrm B} \cdot  \boldsymbol{\mathrm L}) = \tfrac{1}{2} B \cos(\alpha) L_z 
\end{equation}
and the quadratic field term  
\begin{equation}
\label{eqQuadraticMagneticTermAnisotropy}
\boldsymbol{\mathrm A}_{ \rho}^2 = \tfrac{1}{4}{B ^2\rho ^2}\left( {1-\sin ^2(\alpha) \cos ^2 (\phi) } \right)
\end{equation}
are taken into account in the limiting transition from 3D to 2D (see ~\cite{Hansen_2013}).

The problem of the bound states of the 2D hydrogen atom in a magnetic field is described by the Schrodinger equation with the Hamiltonian (\ref{eqRelativeMotionMAIN}) for the relative motion
\begin{equation}
\label{eqBoundStateProblem}
H\Psi({\rho},\phi) =E \Psi({\rho},\phi),
\end{equation}
where $E$ and $\Psi({\rho},\phi)$ are the desired energy level and the wave function of the bound state of relative motion.

In order to find the energy levels $E$ and eigenfunctions $\Psi \left( {\rho ,\phi } \right)$ of the equation (\ref{eqRelativeMotionMAIN}) we use a variation of the discrete variable method, as proposed in \cite{Melezhik_1991} and used in 
\cite{Melezhik_1993} for the investigation of the three-dimensional hydrogen atom in external magnetic and electric fields of arbitrary mutual orientation and in \cite{Koval_2014} for the problem of the dipole-dipole scattering in two spatial dimensions. 
To represent the wave function on a uniform grid ${\phi _j =\frac{2\pi j}{2M+1}}(\mbox{where }j=0,1,...,2M)$ by the angular variable $\phi $, we use eigenfunctions
\begin{equation}
\label{eqKsiDefinition}
\xi _m (\phi )=\frac{1}{\sqrt {2\pi } }e^{im\left( {\phi -\pi } 
\right)}=\frac{(-1)^m}{\sqrt {2\pi } }e^{im\phi },
\end{equation}
of the operator $h^{(0)}(\phi )$ as a Fourier basis. The wave function is sought as the expansion: 
\begin{equation}
\label{eqFullPsiExpansion}
\Psi \left( {\rho ,\phi } \right)=\frac{1}{\sqrt \rho 
} {\sum\limits_{m=-M}^M \sum\limits_{j=0}^{2M}{\xi _m (\phi )\xi 
_{mj}^{-1} \psi _j (\rho )} }, 
\end{equation}
where $\xi _{mj}^{-1} =\frac{2\pi }{2M +1}\xi _{jm}^\ast =\frac{\sqrt{2\pi}
 }{2M +1}e^{-im(\phi _j - \pi) }$ is the inverse matrix to the square matrix $\left({2M +1} \right)\times \left( {2M +1} \right)$ ${\xi _{jm} =\xi _m (\phi _j )}$ determined on the difference grid by the angular variable.
The radial functions $\psi_j(\rho)$ are determined by the values of the wave function on the difference grid  $\phi_j$:
\begin{equation}
\psi_j(\rho)=\sqrt{\rho}\Psi(\rho,\phi_j).
\end{equation}
 
In the representation (\ref{eqFullPsiExpansion}), the Schrodinger equation (\ref{eqBoundStateProblem}) is transformed into a system $2M+1$  of coupled differential equations of the second order:

\begin{multline}
\label{eqShroedEquationInAngularGrid}
\frac{1}{2m_r } 
\left( 
-\frac{\partial^2 }{\partial \rho ^2}\psi_j(\rho)
-\frac{1}{4\rho^2}\psi_j(\rho) +  \sum\limits_{j'=0}^{2M} V_{jj'} \psi_{j'}(\rho) -
\right. 
\\
\left. 
- \frac{1}{\rho^2} \sum\limits_{j'=0}^{2M} h^{(0)}_{jj'} \psi_{j'}(\rho)
\right) = E \psi_j(\rho),
\end{multline}
where the potential matrix has the form: 
\begin{multline}
V_{jj'}(\rho,\phi) = - \frac{2m_r }{\rho}
\delta_{jj'} + (\mu_p-\mu_e) B \cos(\alpha)h^{(1)}_{jj'} + \\
+ \frac{1}{4}B^2\rho^2\left(1-\sin^2(\alpha)\cos^2(\phi_j)\right)\delta_{jj'},
\end{multline}
and the non-diagonal matrix of the operators $h^{(0)}$ and $h^{(1)}\equiv L_z$ are determined by the following ratios: 
\begin{align}
h^{(0)}_{jj'} = - \sum\limits_{j''=-M}^M j''^2 \xi _{jj''} \xi _{j''j'}^{-1}\\
h^{(1)}_{jj'} = \sum\limits_{j''=-M}^M j'' \xi _{jj''} \xi _{j''j'}^{-1}.
\end{align}

The boundary conditions for the radial functions $\psi_j(\rho)$ are determined by the finiteness of the wave function at zero ($\Psi(\rho,\phi_j)=\frac{\psi_j(\rho)}{\sqrt \rho }\to const$)
\begin{equation}
\label{eqLeftBoundaryCondition}
\psi_j(\rho \to 0)\to const \times {\sqrt \rho} \quad (j=0,1,\ldots ,2M )  
\end{equation}
and by its decreasing at infinity:
\begin{equation}
\label{eqRightBoundaryCondition}
\psi_j(\rho \to \infty) \to 0 \quad (j=0,1,\ldots ,2M ).  
\end{equation}

To solve the eigenvalue problem (\ref{eqShroedEquationInAngularGrid}),(\ref{eqLeftBoundaryCondition}),(\ref{eqRightBoundaryCondition}), the nonuniform grid (in the spirit of quasiuniform grids~\cite{Kalitkin_2005}) of the radial variable $\rho$: $\rho_j=\rho_{N} t_{j}^{2}, \quad (j=1,2,\ldots ,N)$ is introduced. Its nodes are determined by mapping $\rho_j \in [0,\rho_{N} \to \infty]$ onto a uniform grid $t_{j}\in [0,1]$.

For discretization we used a finite-difference approximation of the sixth-order accuracy. The eigenvalues of the obtained Hamiltonian matrix are numerically determined by the method of shifted inverse iterations. The algebraic problem arising at each iteration is solved by matrix modification of the sweep algorithm ~\cite{Gelfand_2000} for the block-diagonal matrix.

\section{RESULTS}
\label{Results}

\subsection{Born-Oppenheimer approximation}

Originally, in order to solve the foregoing problem, we applied the Born-Oppenheimer approximation used in many of the papers mentioned in the introduction and dedicated to the 2D hydrogen atom. The Born-Oppenheimer approximation suggests that the electron moves in the field of a positively charged fixed center.

To verify the applied numerical algorithm we studied the problem of the bound states of the 2D hydrogen atom in \textit{the absence of external fields} in the approximation $m_p \to \infty$, having the analytical solution~\cite{Zaslow_1967,Huang_1979,Hassoun_1981,Yang_1991}. 
Table~\ref{tab2DHydrogenSpectrum} demonstrates full agreement of the first ten energy levels of the 2D hydrogen atom, which were calculated numerically using the above-mentioned algorithm, with the analytical values ~\cite{Yang_1991}.
Table~\ref{tabDipoleMatrixElements} shows a comparison of the calculated dipole matrix elements with the ones analytically obtained in~\cite{Yang_1991}.

The algorithm given in Chapter~\ref{sectionAlgorithm} was used to calculate the ground state energy of the 2D hydrogen atom in the magnetic field \textit{perpendicular to the plane of electron motion}. The obtained GSE values that were calculated by us using the infinite proton mass approximation and by other authors using the method of asymptotic iterations~\cite{Soylu_2006} and the variational approach~\cite{Turbiner_2014} for different values of the magnetic field are shown in Table~\ref{tabComparisonWithTurbinerAndSoylu}. The results are consistent with the results of other authors~\cite{Soylu_2006,Turbiner_2014}. The comparison with the values of the energy levels for the two projections of the angular momentum in ~\cite{Taut_1995, Soylu_2006} at different values of the magnetic fields was made. In Table~\ref{tabComparisonWithSoylu_m=0} it is shown for the magnetic quantum number $l=0$ and in Table~\ref{tabComparisonWithSoylu_m=1}~--- for $l=1$.

\subsection{2D hydrogen atom in a magnetic field, considering the finite mass of the proton}
\label{SectionResults_mpLessThanInfty}

For a more accurate analysis of the impact of the magnitude and direction of the magnetic field on the spectrum of the system under study, we calculated the dependence of the 2D hydrogen atom GSE on the magnetic field tilt angle $\alpha$  and on the magnetic field strength, taking into account \textit{the finite mass of the proton}. Figure \ref{fig2} illustrates the dependence found for the field range from $0$ to $4$ a.u. (1~a.u.~=~$2.35~\cdot~10^5$~T) and Figure \ref{figStrongFields}~--- for the field range from  $1$ to  $10^4$ a.u.. Figures~\ref{fig2}--\ref{figStrongFields} illustrate the observed effect of non-linear decrease in the GSE: at low magnetic fields the change is weak, whereas at high magnitudes the GSE is reduced almost by half, with the tilt angle~$\alpha$ changing from~$0$~to~$90^{\circ}$. Note that the found non-linear dependence is not limited only to the impact of the magnetic field projection $B_z$ on the normal to the plane of electron motion, as in this case for the angle $\alpha = 90^{\circ}$, in which $B_z=0$, the ground state energy would be strictly equal to the GSE of the 2D hydrogen atom in the absence of a magnetic field, which is refuted by the results of numerical analysis shown in Fig.~\ref{fig2}. 

In order to estimate the influence of the proton finite mass on the changes in GSE, the calculated GSE values for different magnetic fields and angles $\alpha$ in the $m_p \to \infty$ approximation and the ones, calculated with taking into account the finite mass of the proton, are given in Table \ref{ComparisonBOWithFiniteMass}. It should be highlighted that the results calculated in both approximations differ in the 3rd decimal place in a wide range of the input estimated data.

Analysis of the dependences in Fig.~\ref{fig2} and Fig. \ref{figStrongFields} shows that for non-zero angles $\alpha$ one can observe a slower growth in the GSE values with an increasing magnetic field, although the GSE dependence quadratic for the weak fields, as well as that linear for the strong fields, on the magnetic field remains unchanged. The GSE values for the weak
\begin{equation}
\label{eqTurbinerApproximationForWeakField}
E = -2m_r + \frac{3}{64m_r^3}B^2+\cdots,
\end{equation}
and strong
\begin{equation}
\label{eqTurbinerApproximationForStrongField}
E = \frac{B}{2m_r} - \sqrt{ \frac{\pi}{2} B}+\cdots
\end{equation}
magnetic fields within the perturbation theory that were found by A. V. Turbiner in~\cite{Turbiner_2014}, are represented by the dashed curve in the plane $\alpha=0^{\circ}$ in Fig. \ref{fig2} and Fig. \ref{figStrongFields}, respectively.

The spatial distribution of the wave function of the ground state with an increasing angle  $\alpha$ becomes strongly anisotropic. In particular, one can observe elongation of the atomic state along the axis  $x$ with its simultaneous compression along the axis $y$, which is illustrated in Fig.~\ref{figWaveFunctionProfiles}. Besides the decreasing GSE at the increasing angle $\alpha$ such a behavior of the wave function can be explained by a gradual weakening of the ``harmonic'' contribution of the magnetic field~(\ref{eqQuadraticMagneticTermAnisotropy}) along the axis $x$ at $\alpha \to 90^{\circ}$, which completely disappears in the case of $\alpha = 90^{\circ}$. The corresponding change in the potential surface of the Coulomb and anisotropic quadratic-field (\ref{eqQuadraticMagneticTermAnisotropy}) terms is illustrated in Fig. \ref{figPotentialSurface}.  

\section{CONCLUSION}
\label{Conclusions}

The problem of the 2D hydrogen atom in a magnetic field has been extended for the case of an arbitrary direction of the magnetic field. The algorithm proposed for its solving has been verified using the tasks of the 2D hydrogen atom in the absence of external fields and 2D hydrogen atom in a magnetic field directed perpendicular to the plane of electron motion. For generalization in the case of an arbitrary direction of the magnetic field, the dependence of the ground state energy on the magnetic field and on the tilt angle $\alpha$ between the magnetic field direction and the normal to the plane of electron motion has been calculated. In the process of numerical calculations it has been shown that at an increase in $\alpha$ from  $0^{\circ}$ to $90^{\circ}$ the GSE decreases in a non-linear manner.

The authors express their gratitude to V.S. Melezhik and V.V. Pupyshev for fruitful discussions of this article.

This work was supported by the RFBR (grant No.16-32-00865).


\twocolumn

\nopagebreak

\newpage

\begin{figure}[htbp]
\centerline{\includegraphics[width=0.6 \linewidth]{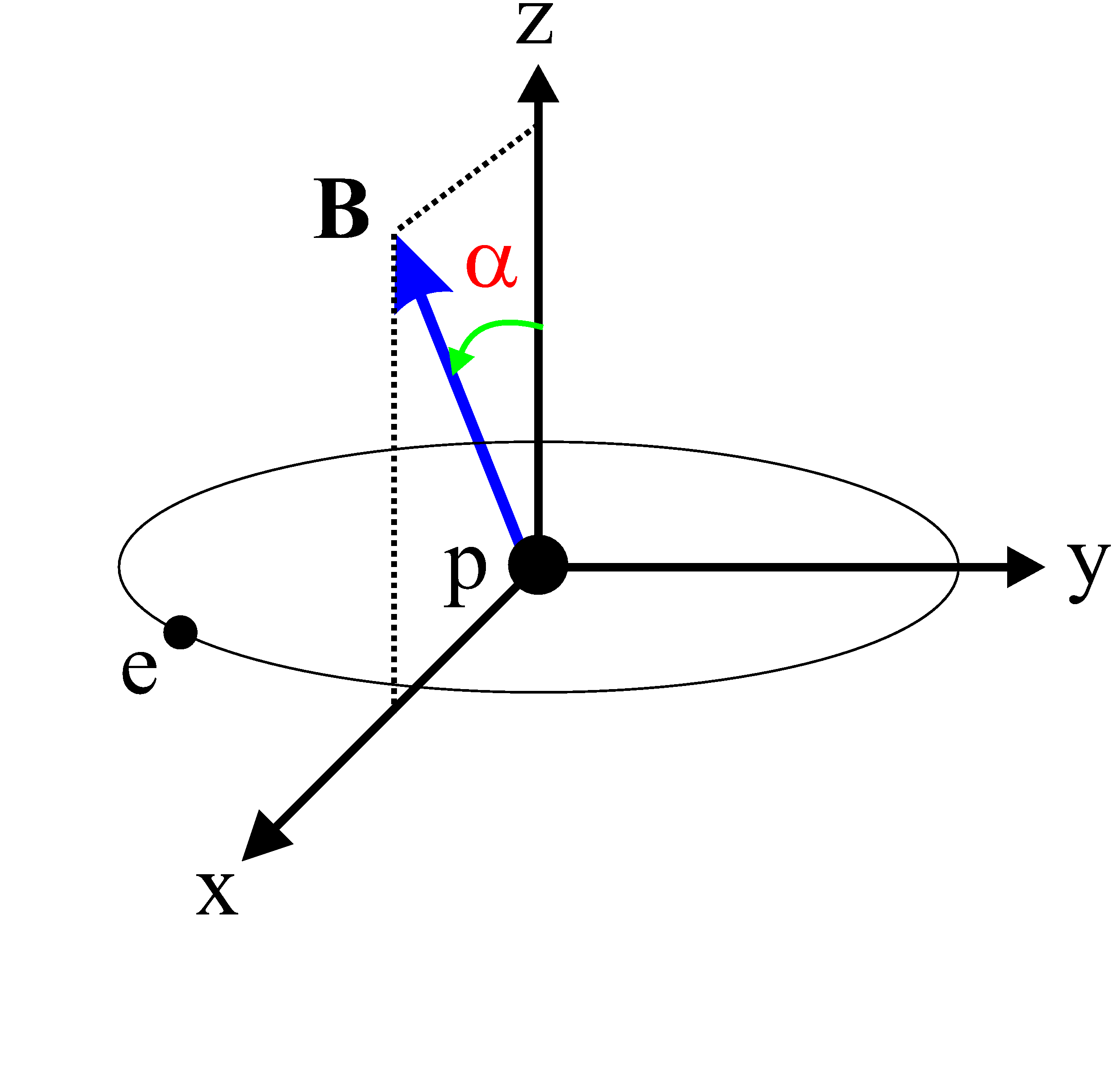}}
\caption{(In color online) Schematic representation of the 2D hydrogen atom in an external magnetic field tilted to an angle $\alpha$ with respect to the normal to the plane of electron motion. 
}
\label{fig1}
\end{figure}

\newpage

\begin{figure}[b]
\centerline{\includegraphics[width= \linewidth]{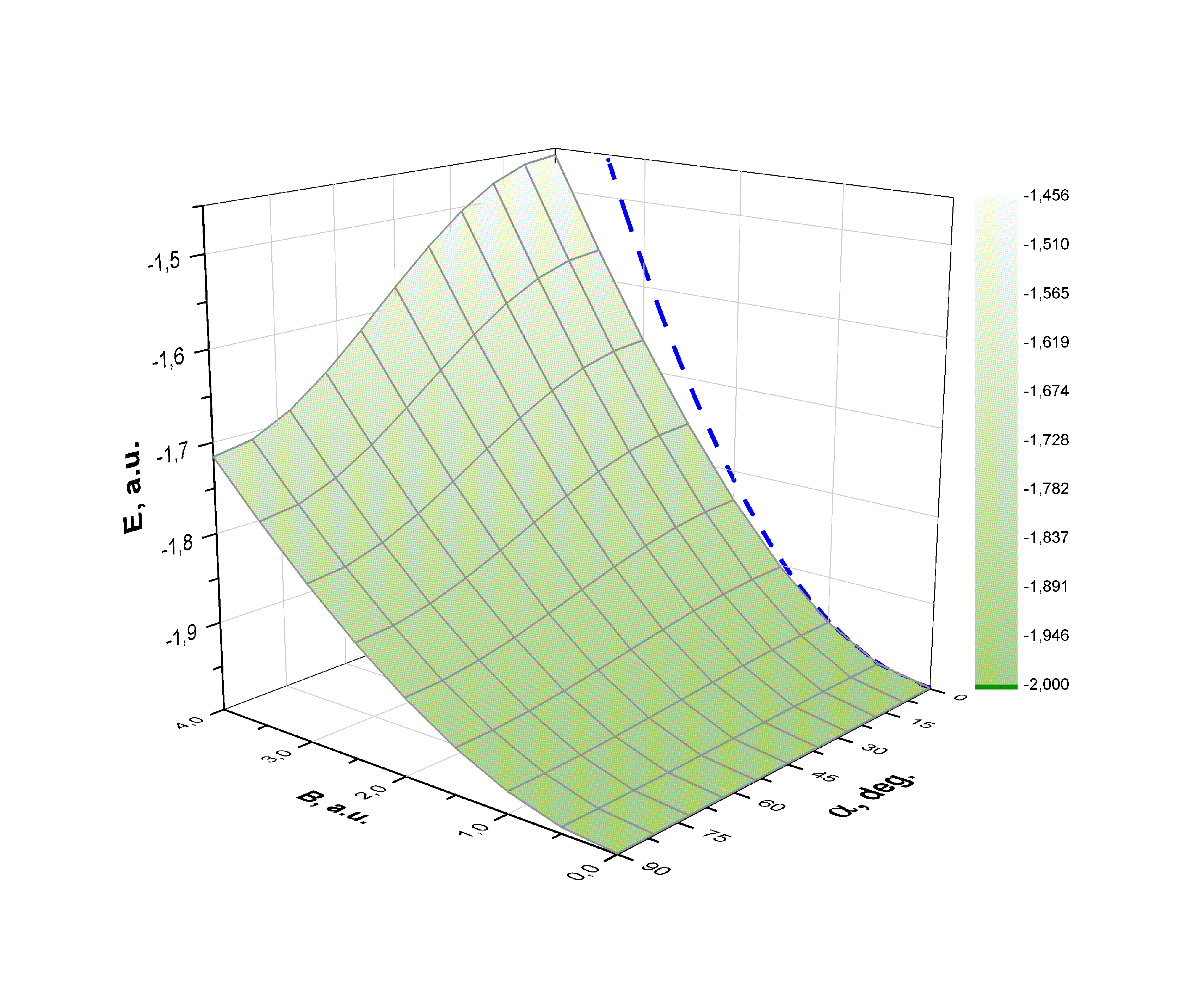}}
\caption{(In color online) Dependence of the 2D hydrogen atom GSE on the magnetic field magnitude and the magnetic field tilt angle $\alpha$, considering the finite proton mass. The dashed curve indicates the GSE dependence for weak magnetic fields  (\ref{eqTurbinerApproximationForWeakField})~\cite{Turbiner_2014}, found according to the perturbation theory.
}
\label{fig2}
\end{figure}

\newpage

\begin{figure}[htbp]
\centerline{\includegraphics[width= \linewidth]{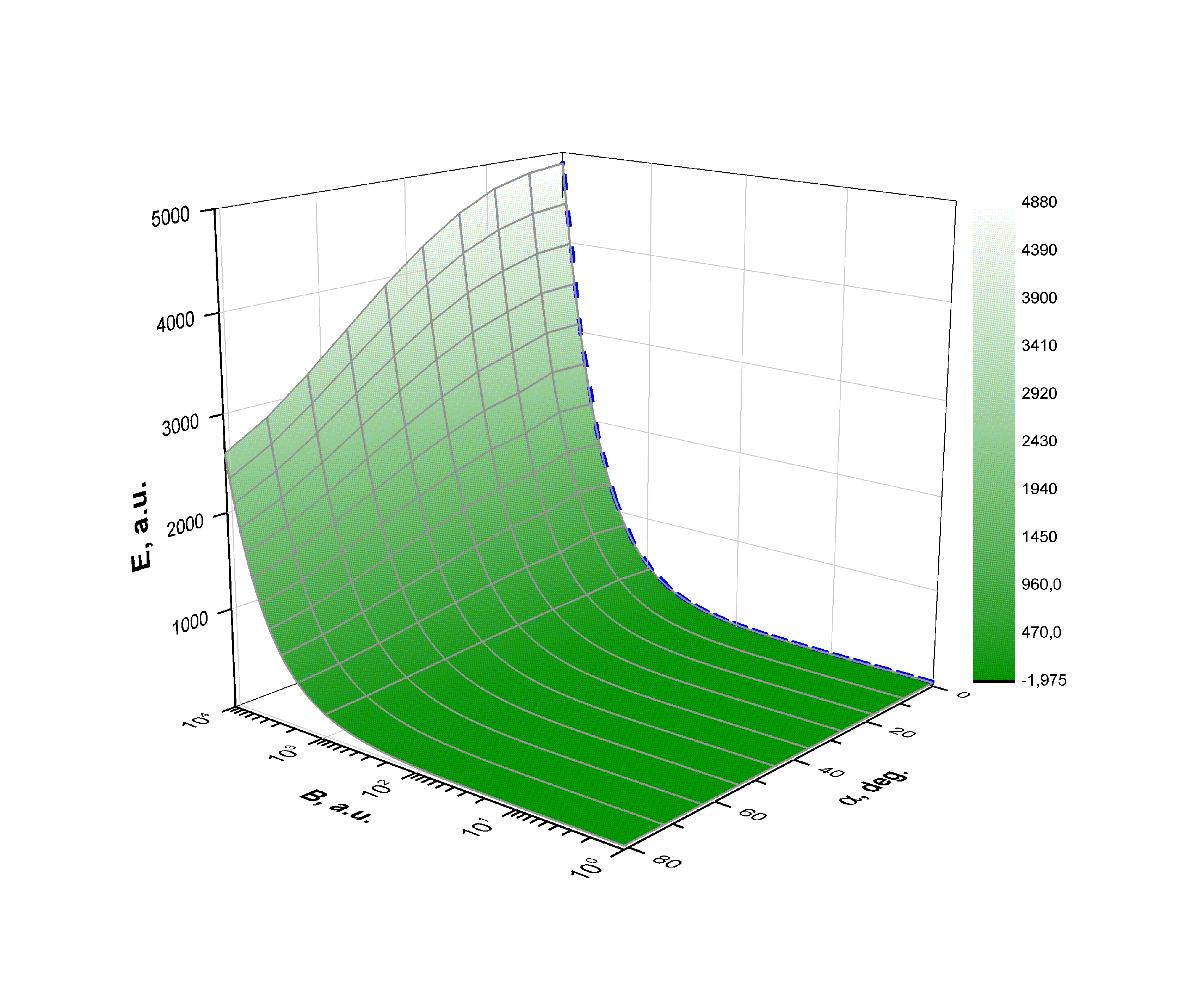}}
\caption{(In color online) Same as that in Fig. \ref{fig2}, for the magnetic field range from  $1$ to $10^4$~a.u.. The dashed curve indicates the GSE dependence for strong magnetic fields  (\ref{eqTurbinerApproximationForStrongField})~\cite{Turbiner_2014}, found according to the perturbation theory.
}
\label{figStrongFields}
\end{figure}

\newpage

\onecolumn

 \begin{figure*}[hbtp]
\begin{minipage}[h]{0.3\linewidth}
\begin{overpic}[height=5cm]{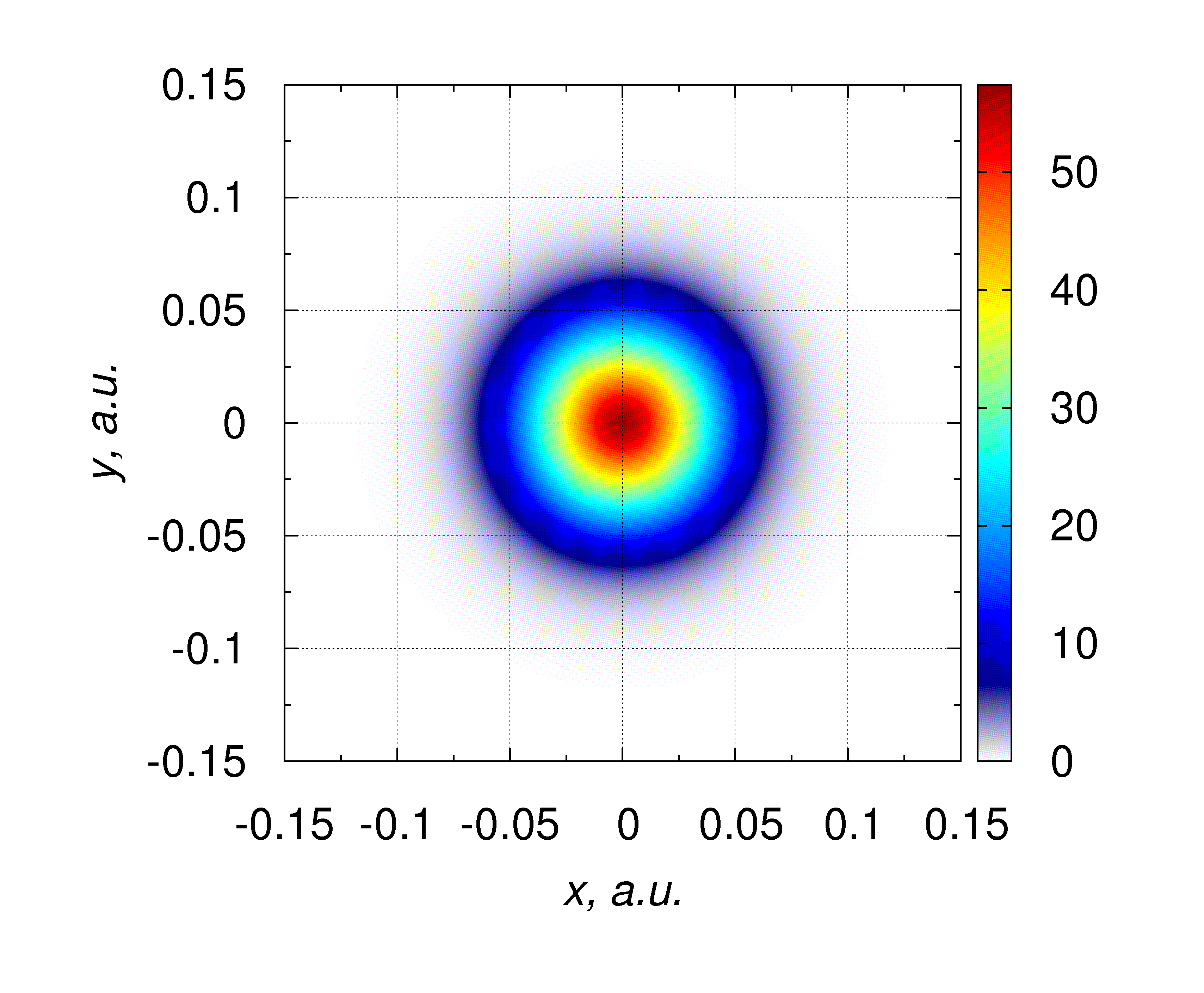}
\put(73,67){\textit{a}}
\end{overpic}
\end{minipage}
\hfill \hfill
\begin{minipage}[h]{0.3\linewidth}
\begin{overpic}[height=5cm]{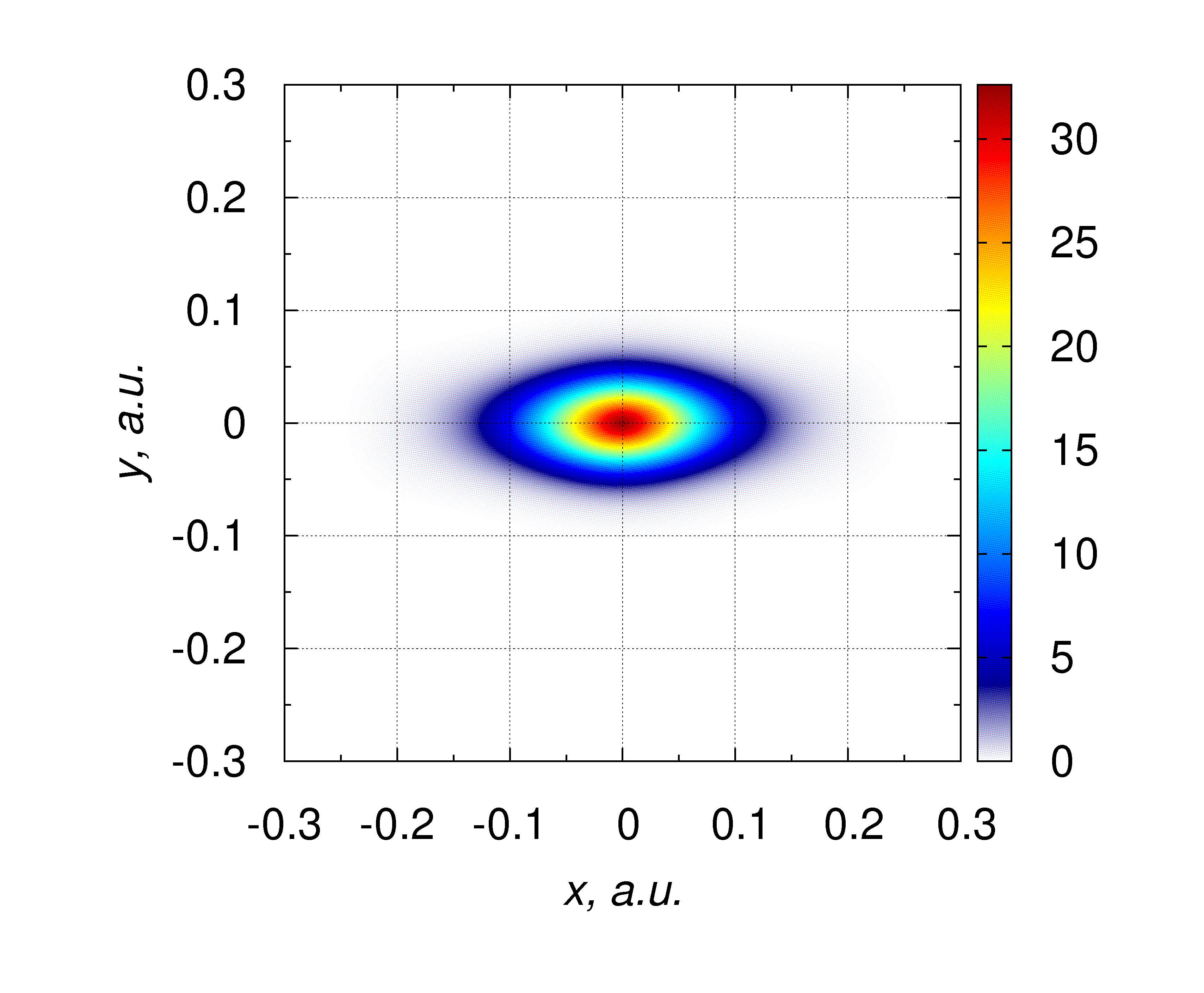}
\put(73,67){\textit{b}}
\end{overpic}
\end{minipage}
\hfill \hfill
\begin{minipage}[h]{0.3\linewidth}
\begin{overpic}[height=5cm]{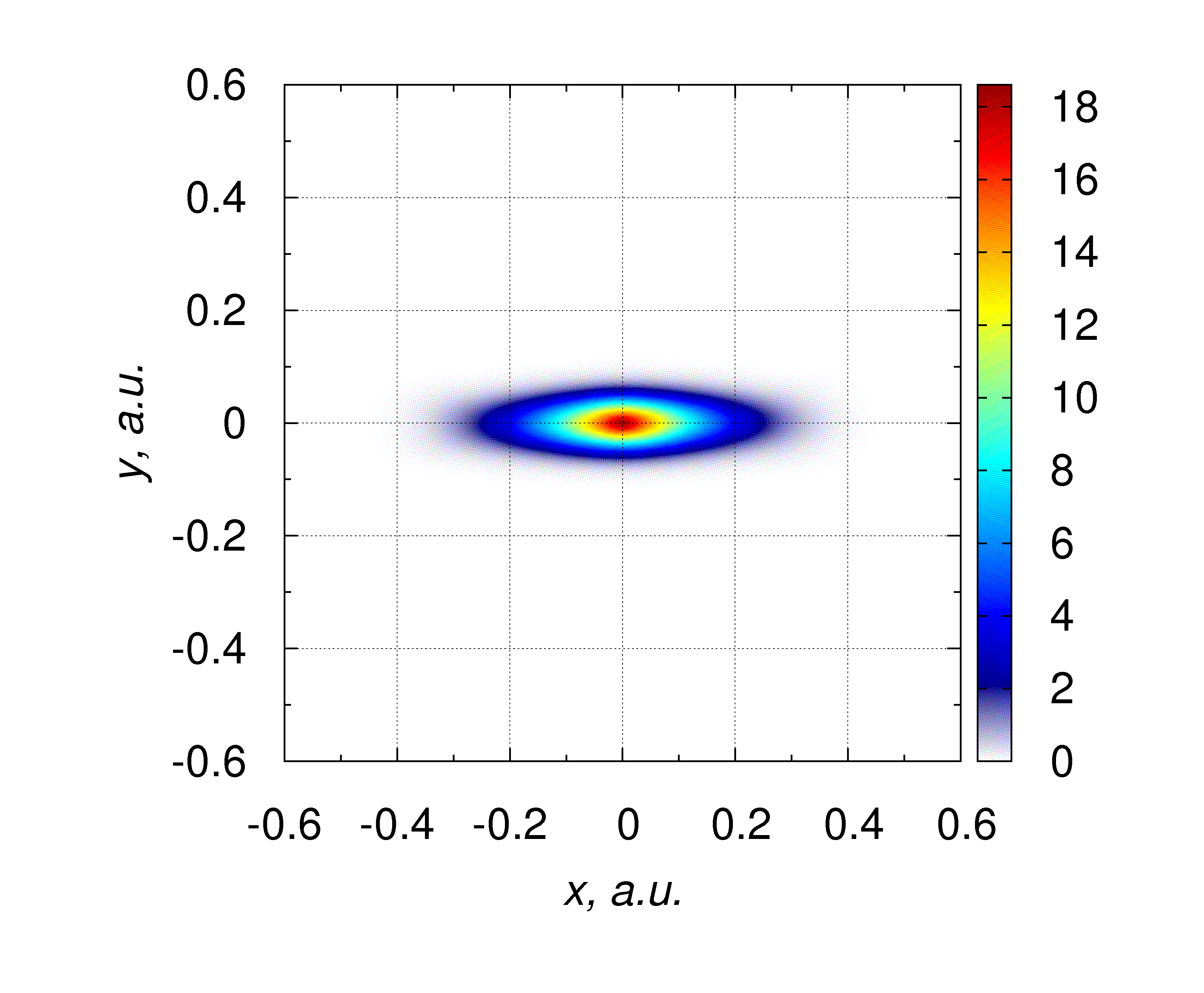}
\put(73,67){\textit{c}}
\end{overpic}
\end{minipage}
\caption{(In color online) Spatial distribution $|\Psi(\boldsymbol \rho)|^2$ of the wave function of the 2D hydrogen atom ground state for the magnetic field $B=10^3 \mbox{ a.u.   }$ at various values of the tilt angle $\alpha$: $\alpha=0^{\circ}$ (\textit{a}),  $\alpha=45^{\circ}$  (\textit{b}),   $\alpha=80^{\circ}$ (\textit{c}).}
 \label{figWaveFunctionProfiles}
 \end{figure*}

\begin{figure*}[hbtp]
\begin{minipage}[h]{0.2\linewidth}
\begin{overpic}[height=5cm]{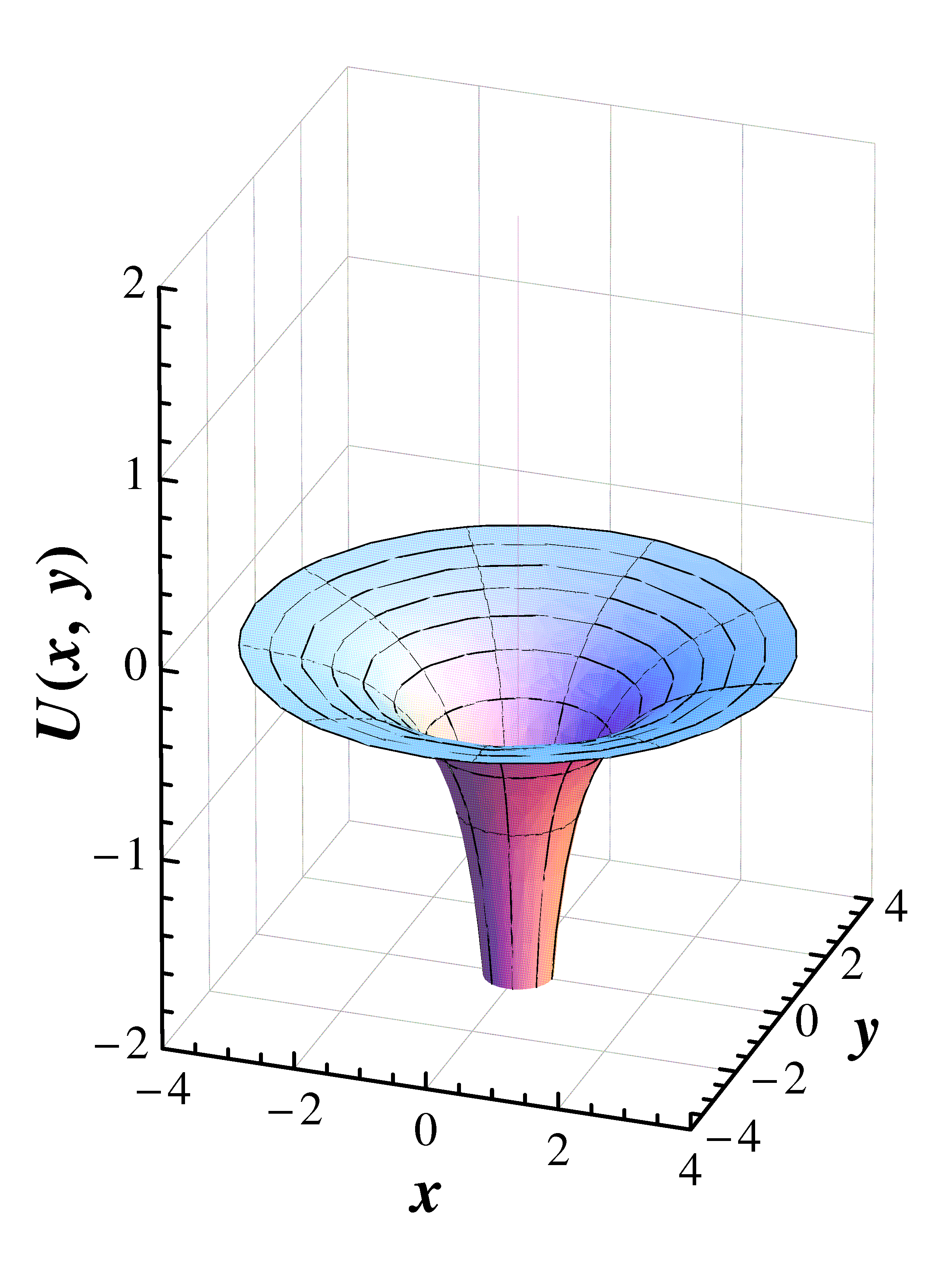}
\put(10,90){\textit{a}}
\end{overpic}
\end{minipage}
\hfill
\begin{minipage}[h]{0.2\linewidth}
\begin{overpic}[height=5cm]{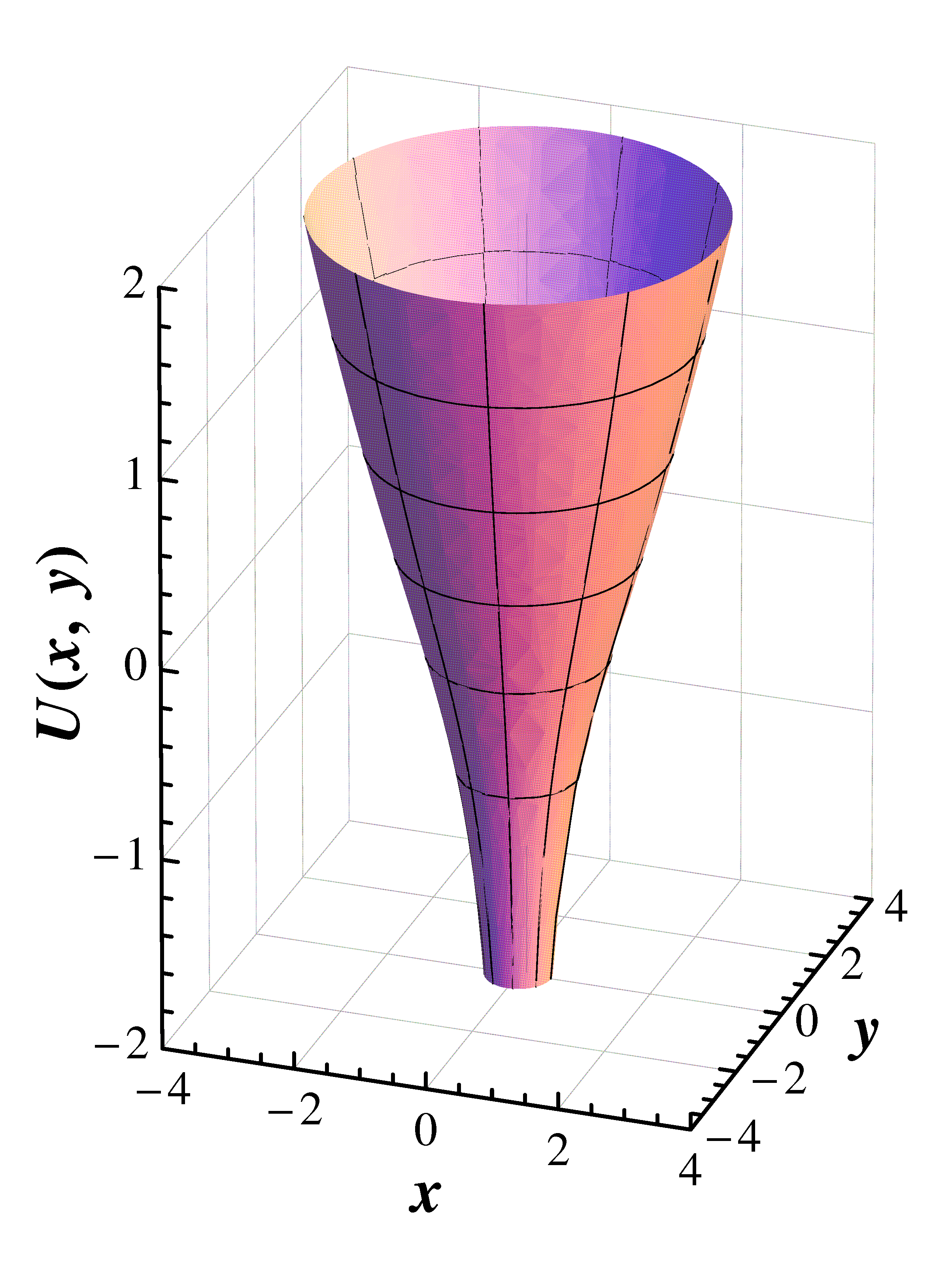}
\put(10,90){\textit{b}}
\end{overpic}
\end{minipage}
\hfill
\begin{minipage}[h]{0.2\linewidth}
\begin{overpic}[height=5cm]{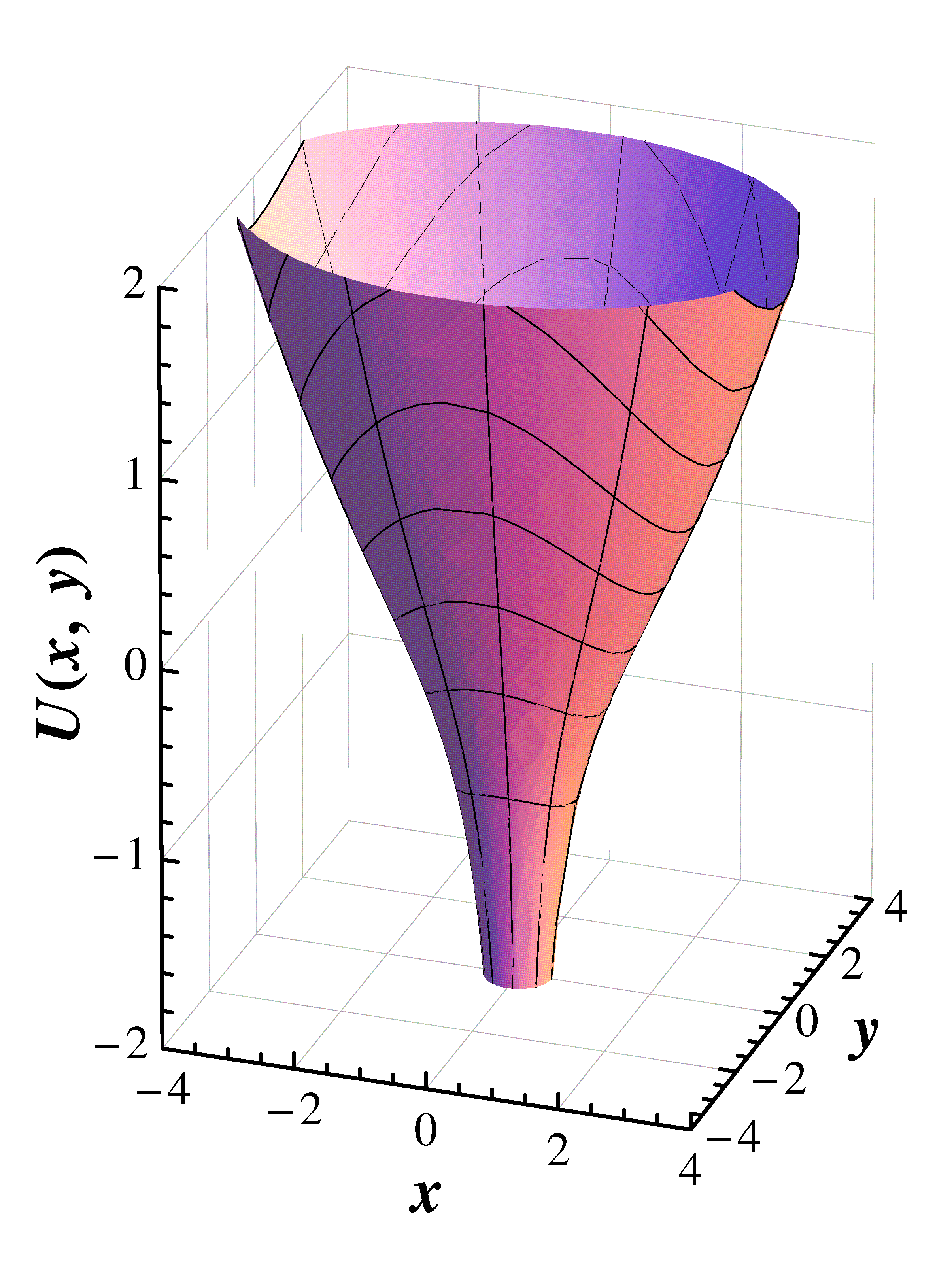}
\put(10,90){\textit{c}}
\end{overpic}
\end{minipage}
\hfill
\begin{minipage}[h]{0.2\linewidth}
\begin{overpic}[height=5cm]{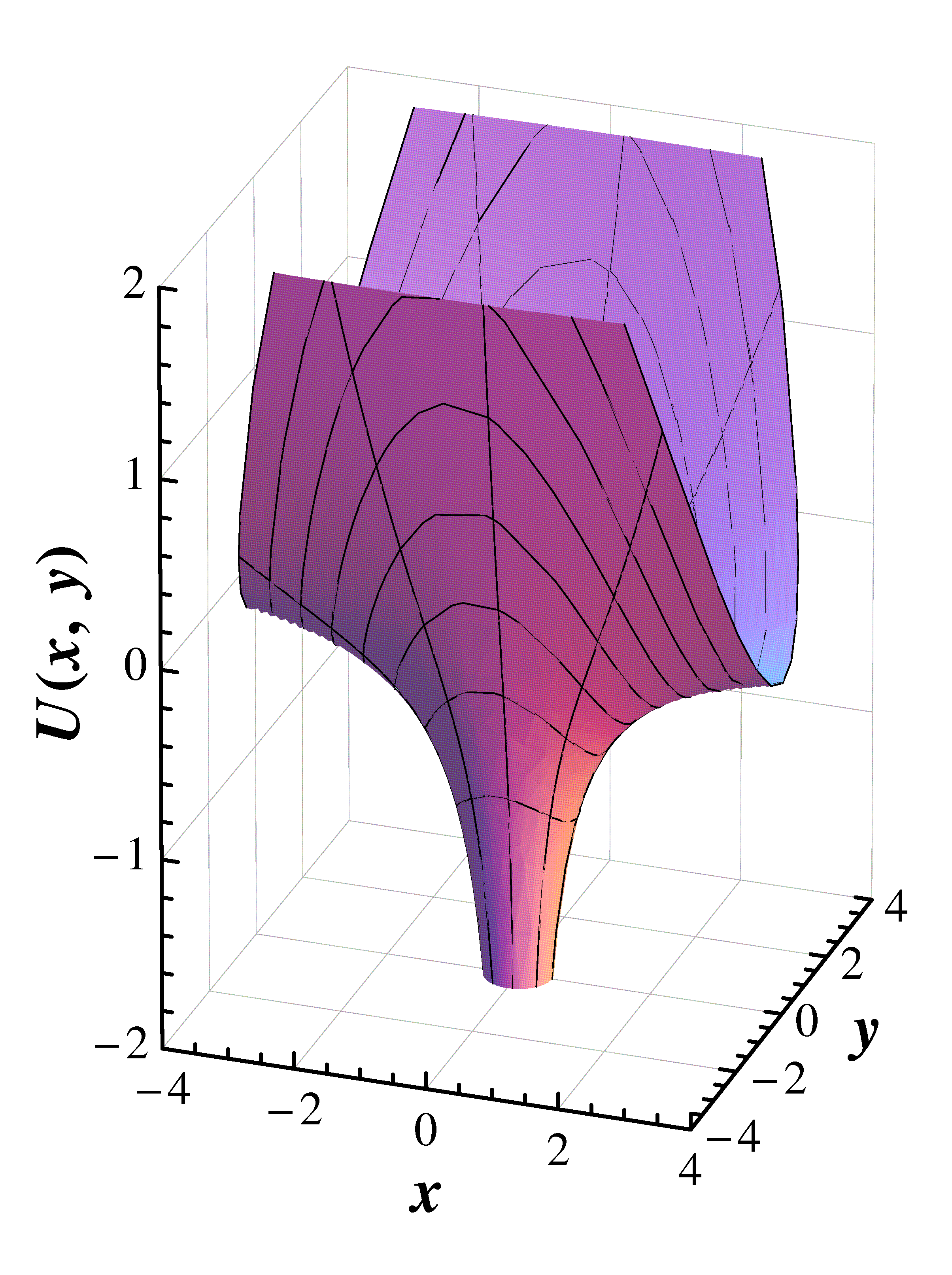}
\put(10,90){\textit{d}}
\end{overpic}
\end{minipage}
\caption{(In color online) Dependence of the potential $U(\boldsymbol{\mathrm \rho}) = -1/\rho + {B ^2\rho ^2}\left( {1-\sin ^2(\alpha) \cos ^2 (\phi) } \right)$ on the magnetic field tilt angle $\alpha$: for the case of magnetic field absence (\textit{a}), for the magnetic field $B=0.5 \mbox{ a.u. }$ at various values of the angle $\alpha$: $\alpha=0^{\circ}$(\textit{b}),   $\alpha=45^{\circ}$ (\textit{c}),  $\alpha=90^{\circ}$  (\textit{d}). All values are given in atomic units.}
 \label{figPotentialSurface}
 \end{figure*}

 \twocolumn

\newpage

\begin{table}
\caption{
\label{tab2DHydrogenSpectrum}
Calculated values of the 2D hydrogen atom energy for the n-th level $E_n$ in the absence of external fields and analytical values $E$ given in~\cite{Yang_1991}.}
\begin{center} 
\begin{tabular}{|c c c|} 
\hline 
\rule{0mm}{5mm}$n$ & $E$, a.u.~\cite{Yang_1991} & $E_{n}$, a.u. \\ 
\hline  
$1$ & $-2.00000000$ & $-2.00000000$ \\ 
$2$ & $-0.22222222$ & $-0.22222222$ \\
$3$ & $-0.08000000$ & $-0.08000000$ \\
$4$ & $-0.04081632$ & $-0.04081633$ \\
$5$ & $-0.02469136$ & $-0.02469136$ \\
$6$ & $-0.01652892$ & $-0.01652892$ \\
$7$ & $-0.01183432$ & $-0.01183432$ \\
$8$ & $-0.00888889$ & $-0.00888889$ \\
$9$ & $-0.00692042$ & $-0.00692042$ \\
$10$ & $-0.00554017$ & $-0.00554016$ \\
\hline 
\end{tabular} 
\end{center} 
\end{table}

\begin{table}
\caption{
\label{tabDipoleMatrixElements}
Comparison of the dipole matrix elements $d_{nl}$, calculated for the 2D hydrogen atom in the absence of external fields and obtained analytically in~\cite{Yang_1991}. 
}
\begin{center} 
\begin{tabular}{|c c cc|} 
\hline 
\rule{0mm}{5mm}$n$ & $l$ &
$d_{nl}$, a.u.~\cite{Yang_1991} & $d_{nl}$, a.u.  \\ 
\hline 
\rule{0mm}{5mm} 
$2$ & $1$ & $0.34445950$ & $0.34445950$ \\ 
$3$ & $1$ & $0.14087514$ & $0.14087514$ \\
$4$ & $1$ & $0.08223128$ & $0.08223128$ \\
$5$ & $1$ & $0.05564053$ & $0.05564053$ \\
\hline 
\end{tabular} 
\end{center} 
\end{table}

\begin{table}
\caption{
\label{tabComparisonWithTurbinerAndSoylu}
GSE values at different magnetic fields obtained by other authors using the method of asymptotic iterations~\cite{Soylu_2006} and variational approach~\cite{Turbiner_2014} compared with the values calculated in the infinite proton mass approximation.    
}
\begin{center} 
\begin{tabular}{|cccc|} 
\hline
\rule{0mm}{5mm} $B$, a.u. & $E$, a.u.~\cite{Soylu_2006} & $E$, a.u.~\cite{Turbiner_2014} & $E$, a.u.  \\ 
\hline 
\rule{0mm}{5mm}  $0.1$ & $-1.999530$ & $-1.999531$ & $-1.999531$ \\ 
$0.25$ & $-1.997078$ & $-1.997079$ & $-1.997079$ \\
$107/250$ & $-1.991490$ & $-1.991490$ & $-1.991491$ \\
$1$ & $-1.955159$ & $-1.955159$ & $-1.955159$ \\
\hline 
\end{tabular} 
\end{center} 
\end{table}

\begin{table}
\caption{
\label{tabComparisonWithSoylu_m=0}
Comparison of the calculated values of the energy levels in the limit $m_p \to \infty$  with the results of~\cite{Soylu_2006,Taut_1995} for the magnetic quantum number  $l=0$.
}
\begin{center} 
\begin{tabular}{|c c ccc|} 
\hline
\rule{0mm}{5mm}$n$ & $B$, a.u. & $E_n$, a.u.~\cite{Taut_1995} & $E_n$, a.u.~\cite{Soylu_2006} & $E_n$, a.u.  \\ 
\hline 
\rule{0mm}{5mm} 
$2$  & $4$ & $4.000000$ & $4.0000000$ & $4.0000000$ \\ 
$3$  & $0.6666666$ & $1.000000$ & $1.0000000$ & $1.0000000$ \\
$4$  & $0.2157031$ & $0.431406$ & $0.4314064$ & $0.4314064$ \\
$3$  & $2.7472602$ & $5.494520$ & $5.4945207$ & $5.4945207$ \\
$5$  & $0.0947113$ & $0.236778$ & $0.2367785$ & $0.2367785$ \\
$4$  & $0.5150444$ & $1.287610$ & $1.2876109$ & $1.2876109$ \\
$6$  & $0.0496114$ & $0.148834$ & $0.1488343$ & $0.1488343$ \\
$5$  & $0.1776672$ & $0.533000$ & $0.5330020$ & $0.5330021$ \\
$4$  & $2.1513889$ & $6.454170$ & $6.4541668$ & $6.4541668$ \\
$10$ & $0.0088435$ & $0.0442176$ & $0.0442177$ & $0.0442177$ \\
\hline 
\end{tabular} 
\end{center} 
\end{table}

\begin{table}
\caption{
\label{tabComparisonWithSoylu_m=1}
Same as in Table \ref{tabComparisonWithSoylu_m=0} for the magnetic quantum number $l=1$.
}
\begin{center} 
\begin{tabular}{| c c ccc|} 
\hline
\rule{0mm}{5mm}$n$ & $B$, a.u. & $E_n$, a.u.~\cite{Taut_1995} & $E_n$, a.u.~\cite{Soylu_2006} & $E_n$, a.u. \\ 
\hline 
\rule{0mm}{5mm} 
$2$ & $1.3333333$ & $2.6666700$ & $2.6666667$ & $2.6666667$ \\ 
$3$ & $0.2857142$ & $0.7142860$ & $0.7142857$ & $0.7142857$ \\
$4$ & $0.1102572$ & $0.3307720$ & $0.3307717$ & $0.3307717$ \\
$3$ & $1.0749278$ & $3.2247800$ & $3.2247835$ & $3.2247835$ \\
$5$ & $0.0545241$ & $0.1909910$ & $0.1907883$ & $0.1907883$ \\
$4$ & $0.2395487$ & $0.8384210$ & $0.8384207$ & $0.8384207$ \\
$6$ & $0.0311049$ & $0.1244200$ & $0.1244197$ & $0.1244197$ \\
$5$ & $0.0951651$ & $0.3806600$ & $0.3806606$ & $0.3806606$ \\
$4$ & $0.9151684$ & $3.6606800$ & $3.6606737$ & $3.6606737$ \\
$7$ & $0.0194448$ & $0.0875018$ & $0.0875018$ & $0.0875018$ \\ 
$10$ & $0.0066281$ & $0.0397691$ & $0.0397691$ & $0.0397691$ \\
\hline 
\end{tabular} 
\end{center} 
\end{table}

\begin{table}
\caption{
\label{ComparisonBOWithFiniteMass}
GSE values calculated for different magnetic fields and angles $\alpha$ in the approximation $m_p \to \infty$ and considering the finite proton mass. 
}
\begin{center} 
\begin{tabular}{|cccc|} 
\hline
\rule{0mm}{5mm}$B$, a.u. & $\alpha$ & $E (m_{p} \to \infty)$, a.u. & $E$, a.u.  \\ 
\hline 
\rule{0mm}{5mm} 
$0$ & $0^{\circ}$ & $-2.00000000$ & $-1.99891136$ \\ 
$1$ & $0^{\circ}$ & $-1.95515969$ & $-1.95400154$  \\
$1$ & $45^{\circ}$ & $-1.96609353$ & $-1.96495184$ \\
$1$ & $90^{\circ}$ & $-1.97736937$  & $-1.97624499$\\
$1.5$ & $0^{\circ}$ & $-1.90335296$ & $-1.90212093$ \\
$1.5$ & $45^{\circ}$ & $-1.92643285$ & $-1.92523340$\\
$1.5$ & $90^{\circ}$ & $-1.95085064$ & $-1.94968360$ \\
$4$ & $0^{\circ}$ & $-1.45958714$ & $-1.45782964$ \\
$4$ & $45^{\circ}$ & $-1.57808514$ & $-1.57646979$ \\
$4$ & $90^{\circ}$ & $-1.71786453$ & $-1.71556932$ \\
\hline 
\end{tabular} 
\end{center} 
\end{table}

}
\end{document}